\documentclass[
 reprint,
 superscriptaddress,
 amsmath,amssymb,
 aps,
 prl,
]{revtex4-2}

\usepackage{enumerate}
\usepackage{booktabs}
\usepackage{dcolumn}
\usepackage{graphicx}
\usepackage{multirow}
\usepackage[colorlinks=true, citecolor=blue, urlcolor=blue, linkcolor=blue]{hyperref}
\usepackage{breakurl}
\usepackage[dvipsnames]{xcolor}

\usepackage{tikz}
\usetikzlibrary{patterns}
\usepackage{pgfplots}
\pgfplotsset{compat=1.5.1}
\usetikzlibrary{arrows.meta,calc,decorations.markings,math,arrows.meta}

\let\originalleft\left
\let\originalright\right
\renewcommand{\left}{\mathopen{}\mathclose\bgroup\originalleft}
\renewcommand{\right}{\aftergroup\egroup\originalright}

\newcommand{\bunderbrace}[2]{%
  \begin{array}[t]{@{}c@{}}
  \underbrace{#1}\\
  #2
  \end{array}
}

\begin{document}

%TC:ignore

\title{Local-Order Fluctuations in Kob-Andersen-Type Glass Formers}

\author{John \c{C}amk{\i}ran}
\email{john.camkiran@utoronto.ca}
\affiliation{\mbox{Department of Materials Science and Engineering, University of Toronto, Toronto, Ontario, M5S 3E4, Canada}}

\author{Fabian Parsch}
\affiliation{\mbox{Department of Mathematics, University of Toronto,  Toronto, Ontario, M5S 2E4, Canada}
}
\affiliation{\mbox{Department of Materials Science and Engineering, University of Toronto, Toronto, Ontario,  M5S 3E4, Canada}}

\author{Glenn D. Hibbard}
\affiliation{\mbox{Department of Materials Science and Engineering, University of Toronto, Toronto, Ontario, M5S 3E4, Canada}}

\date{\today}

\begin{abstract}
Notwithstanding decades of work, we still lack a satisfactory understanding of the structural relaxation that takes place as a liquid is rapidly cooled to form a glass. The present paper discusses a novel statistical characterization of this phenomenon in Kob-Andersen-type mixtures---a simple yet powerful class of model glass formers. We use the variance of an order parameter called the extracopularity coefficient to measure the intensity of instantaneous fluctuations in local orientational order. This intensity is found to be nearly composition independent at the onset temperature of glassy dynamics for the standard Kob-Andersen mixture. We decompose these fluctuations into a density and symmetry contribution. Through the behavior of these contributions, we argue that the near composition independence prevails when the structure of the system equally resembles a liquid and a glass. We moreover report that the intensity of local-order fluctuations behaves like a heuristic measure of glass-forming ability, which could prove useful when exact methods are intractable.
\end{abstract}

\maketitle

%TC:endignore

In its almost three decades of existence, the Kob-Andersen (KA) binary mixture of Lennard-Jones fluids \cite{kob_1995} has become a canonical tool for computational research on glass formation. Despite its relative simplicity, the mixture exhibits nearly the entire spectrum of nontrivial behaviors observed in the more complex, molecular glass formers \cite{berthier_2022}, %\cite{turci_2019, gonzalezlopes_2021, mehri_2021}
making it a valuable testing ground for new ideas. Originally devised to model the $80$:$20$ nickel-phosphorus alloy, its strong intercomponent attraction inhibits compositional fluctuations, thereby precluding crystallization via phase separation \cite{toxvaerd_2009}. Studies involving very long simulations have nevertheless succeeded in crystallizing such mixtures, culminating in a structural phase diagram \cite{chen_2018, pedersen_2018}. Such diagrams bear important insight on the fully relaxed, equilibrium states of a system but not on the nature of the structural relaxation that unfolds far from equilibrium. The latter is of particular interest in investigating the possible structural origins of the glass transition \cite{schoenholz_2016, tong_2018, bapst_2020, nandi_2021}, which remain poorly understood \cite{biroli_2013, guiselin_2022}. 

The present work investigates short-time structural relaxation in KA-type mixtures through instantaneous fluctuations in local orientational order, which have recently been shown to play a decisive role in nucleation \cite{becker_2022,hu_2022b}. Here, these fluctuations are found to suggest a classification of such mixtures that is consistent with the main features of their structural phase diagram. Our investigations reveal the presence of a characteristic temperature at which the intensity of local order fluctuations is nearly composition independent. This temperature appears to coincide approximately with the onset temperature of glassy dynamics for the standard KA mixture \cite{banerjee_2017}, pointing to a structure--dynamics relationship hitherto unexplored. 

The glass-forming ability of a liquid is a property of both practical and theoretical interest, conventionally evaluated through the critical cooling rate \cite{pedersen_2018, hu_2022}---the fastest cooling rate for which a system evidences crystallization. But this rate can be difficult to measure, as superior glass formers are precisely those systems that escape crystallization at experimental and computational time scales. Here, we observe that better KA-type glass formers have larger fluctuations in local orientational order, thus establishing order fluctuation intensity as a possible structural heuristic on glass-forming ability. This heuristic suggests that the superlative KA-type glass former has a composition ratio of approximately $70$:$30$. Such is, to the best of our knowledge, the first purely structural piece of evidence supporting the common assertion that the best KA-type glass former has a higher impurity content than the eutectic composition of $74$:$26$
\cite{bruning_2008,pedersen_2018,mehri_2021}, which is the traditional candidate for that designation \cite{turnbull_1969}. Given its lack of explicit assumptions, the method of the present work may provide a tractable, alternative means of evaluating glass-forming ability in other classes of particle-based systems.

\paragraph{Methods} Liquids and glasses can be distinguished by both their microscopic dynamics and their macroscopic properties. However, attempts to do so through an understanding of amorphous structure have found little success \cite{berthier_2010,berthier_2011a,berthier_2011b}. One of the difficulties encountered on the computational end of these efforts is the lack of a reliable way to quantify structure \cite{tanaka_2019, pedersen_2021, nandi_2021}. Motivated thus, the authors of this paper recently introduced a simple local orientational order parameter called the \textit{extracopularity coefficient} \cite{camkiran_2022a}, defined by
\begin{equation}
    E = \log_2  \left[\frac{k^2 - k}{ 2 m } \right], \quad k>1,
\end{equation}
where $k$ denotes the coordination number of the particle under study, and $m$ denotes its number of distinct bond angles. A large $E$ tells us that the geometry of the immediate surrounding of a particle is simple in the sense of lacking diversity in bond angles. For more on $E$, see Ref. \cite{camkiran_2022a} and \cite{camkiran_2022b}.

\begin{figure}[t]
    \centering
    \includegraphics[width=0.9\linewidth]{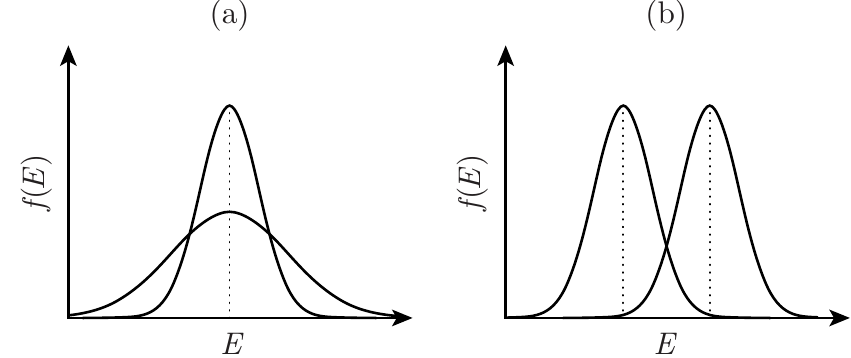}
    \caption{(a) $E$ distributions equal in mean, differing in variance and (b) $E$ distributions equal in variance, differing in mean.}
    \label{fig:schematic}
\end{figure}

Whereas $E$ indicates the instantaneous structure around an individual particle in a system, this work concerns that of a system as a whole. We can obtain a global structural understanding from the local one provided by $E$ by considering its mean $\langle E \rangle$ and variance $\mathrm{var}(E) = \langle E^2 \rangle - \langle E \rangle^2$, which are depicted schematically in Fig. \ref{fig:schematic}.

The variance of $E$ indicates the total intensity of \textit{local-order fluctuations} (LOF). It is possible to decompose this quantity into a density and a symmetry contribution. Observe that $E$ can be written as the sum of a term depending only on $k$ and one depending only on $m$,
\begin{equation}
    E = \bunderbrace{\log_2 \left[ \frac{k^2-k}{2} \right]}{E_k} + \bunderbrace{\log_2 \left[ \frac{1}{m} \right]}{E_m}.
\end{equation}
Taking the variance of both sides, we have
\begin{align}
    \underbrace{\strut \mathrm{var}(E)}_{\text{LOF}} = \underbrace{\strut \mathrm{var}(E_k)}_{\text{LDF}} + \underbrace{\strut \mathrm{var}(E_m) + 2\mathrm{cov}(E_k,E_m)}_{\text{LSF}}.
    \label{eq:decomposition}
\end{align}
Since $E_k$ is strictly monotonic in $k$, the variance of $E_k$ fully captures fluctuations in coordination number, here termed \textit{local-density fluctuations} (LDF). Any remaining variation in $E$ must be accounted for by differences in the ways neighbors are arranged, here termed \textit{local-symmetry fluctuations} (LSF).

We used LAMMPS \cite{thompson_2022} to simulate glass formation in KA-type binary mixtures. Such mixtures consist of a primary component $A$ of larger particles and a secondary component $B$ of smaller particles. Their composition is alternatively specified by the $A$:$B$ composition ratio or the $B$ molar fraction $\chi_B$. The two components have identical masses, $m_A = m_B = 1$, and interact via the Lennard-Jones potential with parameters $\sigma_{AA} = 1$, $\sigma_{AB} = 0.8$, $\sigma_{BB} = 0.88$, $\varepsilon_{AA} = 1$, $\varepsilon_{AB} = 1.5$, and $\varepsilon_{BB} = 0.5$, truncated at $2.5\sigma$. Each of our simulations featured a system of $N=64000$ particles in a cubic box with periodic boundary conditions and was performed using the Nos\'{e}-Hoover thermostat with a timestep size of $\delta t = 0.005$. The systems were first equilibrated in an isothermal-isobaric (NPT) ensemble at a pressure of $P = 10.19$ and a temperature of $T_1 = 2$. The pressure $P$ was chosen to correspond to the available phase diagram \cite{chen_2018, pedersen_2018}, and the temperature $T_1$ was chosen to be approximately double the melting temperature $T_\text{m} = 1.028$ of the $80$:$20$ mixture at that pressure \cite{pedersen_2018}. The systems were then cooled under constant pressure to a temperature of $T_2 = 0.0002 \approx 0$ with rates $R$, $10R$ and $100R$, where the baseline rate $R = \delta T / \delta t = 0.008$ was chosen to be an order of magnitude faster than the fastest rate needed to obtain a single crystal from a $100$:$0$ mixture of $8000$ particles. The damping parameter values we used to hold cooling rate and pressure constant are given in Table \ref{tab:critical-values}.

Extracopularity coefficients were computed using version 0.1.0 of an implementation \footnote{Code available at \url{www.github.com/johncamkiran/extracopularity}} of the algorithm described in Ref. \cite{camkiran_2022a}. For particles without commonly encountered coordination geometries \cite{camkiran_2022b}, this version of the algorithm computes bond angle count $m$ iteratively as follows: For every iteration $i$, bond angles are discretized using bins of random width, each independently sampled from an exponential distribution with mean $7.5^\circ$, truncated below by $5^\circ$. The number of bins filled is assigned to variable $m_i$. Then, $m$ is taken to be the first percentile of $m_i$ over a large number of iterations (here $400$). The mean and lower cutoff of the exponential were chosen to obtain $E > 3$ for close-packed particles, which is consistent with the values of $E$ for FCC and HCP \cite{camkiran_2022a}.

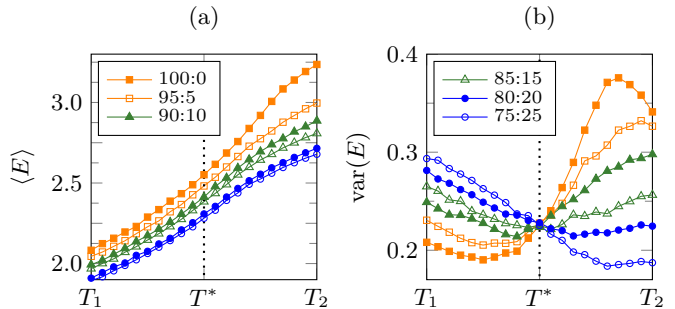
\begin{figure}[b]
\hspace{-4mm}
    \begin{tabular}{cc}
        \begin{tabular}{l}
        	\begin{tikzpicture}[baseline,trim axis right]
            	\begin{axis}[
        	    title = (a),
        		width=0.53\linewidth,
                height=0.53\linewidth,
                ylabel = $\langle E \rangle$,
                ylabel near ticks,
                xmin = 1,
                xmax = 21,
                ymin = 1.90, 
                ymax = 3.30,
                xtick={1, 11, 21},
                xticklabels = {$T_\textrm{1}$, $T^*$,$T_\textrm{2}$},
                ytick = {1.75, 2.0, 2.25, 2.5, 2.75, 3.00, 3.25},
                yticklabels = {, 2.0, , 2.5, , 3.0},
                scaled ticks = false,
                legend pos = north west,
                legend style={font=\scriptsize,row sep=-0.1cm,/tikz/every odd column/.append style={column sep=0.01cm}},
                legend cell align={left},
                minor y tick num = 1,
                ]
                \addplot[
                    color=orange,
                    mark=square*,
                    mark size=1.2pt,
                    ]
                coordinates {
                    (1,2.0825)
                    (2,2.1236)
                    (3,2.1573)
                    (4,2.1958)
                    (5,2.2304)
                    (6,2.2885)
                    (7,2.3349)
                    (8,2.3875)
                    (9,2.4339)
                    (10,2.4892)
                    (11,2.5515)
                    (12,2.616)
                    (13,2.6862)
                    (14,2.7575)
                    (15,2.8387)
                    (16,2.9196)
                    (17,3.0066)
                    (18,3.084)
                    (19,3.1392)
                    (20,3.1921)
                    (21,3.2359)
                };
                \addplot[
                    color=orange,
                    mark=square,
                    mark size=1.2pt,
                    ]
                coordinates {
                    (1,2.045)
                    (2,2.0716)
                    (3,2.1069)
                    (4,2.1416)
                    (5,2.1855)
                    (6,2.2283)
                    (7,2.2649)
                    (8,2.3214)
                    (9,2.3661)
                    (10,2.4268)
                    (11,2.4817)
                    (12,2.536)
                    (13,2.5994)
                    (14,2.6644)
                    (15,2.7285)
                    (16,2.7752)
                    (17,2.8232)
                    (18,2.8775)
                    (19,2.9192)
                    (20,2.9606)
                    (21,2.9964)
                };
                \addplot[
                    color=OliveGreen,
                    mark=triangle*,
                    mark size=1.8pt,
                ]
                coordinates {
                    (1,1.9934)
                    (2,2.0203)
                    (3,2.0578)
                    (4,2.0942)
                    (5,2.1366)
                    (6,2.174)
                    (7,2.2141)
                    (8,2.2574)
                    (9,2.3073)
                    (10,2.3592)
                    (11,2.4127)
                    (12,2.4788)
                    (13,2.5322)
                    (14,2.5934)
                    (15,2.646)
                    (16,2.6952)
                    (17,2.7387)
                    (18,2.7762)
                    (19,2.8182)
                    (20,2.8501)
                    (21,2.8865)
                };
                \addplot[
                    color=OliveGreen,
                    mark=triangle,
                    mark size=1.8pt,
                ]
                coordinates {
                    (1,1.9683)
                    (2,2.0006)
                    (3,2.0294)
                    (4,2.0705)
                    (5,2.1049)
                    (6,2.1463)
                    (7,2.1857)
                    (8,2.2298)
                    (9,2.275)
                    (10,2.3296)
                    (11,2.3844)
                    (12,2.4358)
                    (13,2.4905)
                    (14,2.5513)
                    (15,2.5982)
                    (16,2.6432)
                    (17,2.6786)
                    (18,2.7182)
                    (19,2.7501)
                    (20,2.7822)
                    (21,2.8079)
                };
                \addplot[
                    color=blue,
                    mark=*,
                    mark size=1.2pt,
                ]
                coordinates {
                    (1,1.9095)
                    (2,1.9439)
                    (3,1.9793)
                    (4,2.0037)
                    (5,2.051)
                    (6,2.076)
                    (7,2.1218)
                    (8,2.1566)
                    (9,2.209)
                    (10,2.2537)
                    (11,2.308)
                    (12,2.355)
                    (13,2.413)
                    (14,2.4664)
                    (15,2.5142)
                    (16,2.5547)
                    (17,2.5945)
                    (18,2.6265)
                    (19,2.659)
                    (20,2.6877)
                    (21,2.7155)
                };
                \addplot[
                    color=blue,
                    mark=o,
                    mark size=1.2pt,
                ]
                coordinates {
                    (1,1.8943)
                    (2,1.9178)
                    (3,1.9549)
                    (4,1.9887)
                    (5,2.0234)
                    (6,2.0628)
                    (7,2.1015)
                    (8,2.1428)
                    (9,2.1862)
                    (10,2.2282)
                    (11,2.2816)
                    (12,2.3312)
                    (13,2.3924)
                    (14,2.4385)
                    (15,2.4922)
                    (16,2.5247)
                    (17,2.5608)
                    (18,2.5964)
                    (19,2.6259)
                    (20,2.6543)
                    (21,2.6786)
                };
                \legend{$100$:$0$,$95$:$5$,$90$:$10$}
                \addplot[thick, dotted, samples=2, smooth,black] coordinates {(11,0)(11,3.2)};
        	\end{axis}
        \end{tikzpicture}
	\end{tabular}
    &
	\begin{tabular}{l}
    	\begin{tikzpicture}[baseline,trim axis right]
    	    \begin{axis}[
        	    title = (b),
        		width=0.53\linewidth,
                height=0.53\linewidth,
                xmin = 1,
                xmax = 21,
                ymin = 0.17, 
                ymax = 0.40,
                ylabel={$\mathrm{var}(E)$},
                ylabel near ticks,
                xtick={1, 11, 21},
                xticklabels = {$T_\textrm{1}$, $T^*$,$T_\textrm{2}$},
                yminorticks=false,
                scaled ticks = false,
                legend pos = north west,
                legend style={font=\scriptsize,row sep=-0.1cm,/tikz/every odd column/.append style={column sep=0.01cm}},
                legend cell align={left},
                minor y tick num = 1,
                ]
                \addplot[
                    color=orange,
                    mark=square*,
                    mark size=1.2pt,
                ]
                coordinates {
                    (1,0.2081)
                    (2,0.2036)
                    (3,0.1989)
                    (4,0.195)
                    (5,0.1931)
                    (6,0.1902)
                    (7,0.1931)
                    (8,0.1972)
                    (9,0.1991)
                    (10,0.2123)
                    (11,0.2252)
                    (12,0.2403)
                    (13,0.2639)
                    (14,0.2896)
                    (15,0.3225)
                    (16,0.3479)
                    (17,0.3712)
                    (18,0.3758)
                    (19,0.3689)
                    (20,0.3582)
                    (21,0.3411)
                };
                \addplot[
                    color=orange,
                    mark=square,
                    mark size=1.2pt,
                    ]
                coordinates {
                    (1,0.2306)
                    (2,0.2244)
                    (3,0.2184)
                    (4,0.2122)
                    (5,0.2078)
                    (6,0.2053)
                    (7,0.2073)
                    (8,0.2071)
                    (9,0.2088)
                    (10,0.2188)
                    (11,0.2277)
                    (12,0.2359)
                    (13,0.2478)
                    (14,0.2663)
                    (15,0.2911)
                    (16,0.2962)
                    (17,0.3072)
                    (18,0.3225)
                    (19,0.3264)
                    (20,0.3321)
                    (21,0.3265)
                };
                \addplot[
                    color=OliveGreen,
                    mark=triangle*,
                    mark size=1.8pt,
                    ]
                coordinates {
                    (1,0.249)
                    (2,0.2424)
                    (3,0.2362)
                    (4,0.2358)
                    (5,0.2321)
                    (6,0.2276)
                    (7,0.2216)
                    (8,0.2161)
                    (9,0.214)
                    (10,0.2215)
                    (11,0.2236)
                    (12,0.2352)
                    (13,0.2448)
                    (14,0.2519)
                    (15,0.2611)
                    (16,0.2728)
                    (17,0.2808)
                    (18,0.2851)
                    (19,0.2922)
                    (20,0.2938)
                    (21,0.2978)
                };
                \addplot[
                    color=OliveGreen,
                    mark=triangle,
                    mark size=1.8pt,
                    ]
                coordinates {
                    (1,0.265)
                    (2,0.2604)
                    (3,0.2514)
                    (4,0.25)
                    (5,0.2398)
                    (6,0.2361)
                    (7,0.2318)
                    (8,0.2274)
                    (9,0.2254)
                    (10,0.2233)
                    (11,0.2255)
                    (12,0.2215)
                    (13,0.2263)
                    (14,0.2361)
                    (15,0.2348)
                    (16,0.239)
                    (17,0.2373)
                    (18,0.2448)
                    (19,0.2498)
                    (20,0.255)
                    (21,0.2564)
                };
                \addplot[
                    color=blue,
                    mark=*,
                    mark size=1.2pt,
                    ]
                coordinates {
                    (1,0.2814)
                    (2,0.2727)
                    (3,0.2683)
                    (4,0.2619)
                    (5,0.2566)
                    (6,0.249)
                    (7,0.2431)
                    (8,0.2369)
                    (9,0.2358)
                    (10,0.2304)
                    (11,0.228)
                    (12,0.222)
                    (13,0.22)
                    (14,0.2146)
                    (15,0.2164)
                    (16,0.2183)
                    (17,0.2177)
                    (18,0.2206)
                    (19,0.2224)
                    (20,0.2258)
                    (21,0.2246)
                };
                \addplot[
                    color=blue,
                    mark=o,
                    mark size=1.2pt,
                    ]
                coordinates {
                    (1,0.2935)
                    (2,0.2918)
                    (3,0.2833)
                    (4,0.2777)
                    (5,0.2712)
                    (6,0.2628)
                    (7,0.2562)
                    (8,0.2486)
                    (9,0.2371)
                    (10,0.2353)
                    (11,0.2254)
                    (12,0.2164)
                    (13,0.2101)
                    (14,0.1982)
                    (15,0.1955)
                    (16,0.1889)
                    (17,0.1838)
                    (18,0.1854)
                    (19,0.1866)
                    (20,0.1885)
                    (21,0.1873)
                };
                \addplot[thick, dotted, samples=2, smooth,black] coordinates {(11,0.00)(11,0.40)};
                \legend{,,,$85$:$15$, $80$:$20$, $75$:$25$}
            \end{axis}
    	\end{tikzpicture}
	\end{tabular}
\end{tabular}
\caption{Statistics of $E$ versus (decreasing) temperature for hypoeutectic KA-type mixtures: (a) mean and (b) variance.}
\makeatletter
\let\save@currentlabel\@currentlabel
\edef\@currentlabel{\save@currentlabel(a)}\label{fig:statistics-mean}
\edef\@currentlabel{\save@currentlabel(b)}\label{fig:statistics-var}
\end{figure}

\paragraph{Results} As a first step towards understanding how the structure of KA-type  mixtures change when rapidly cooled, we considered the average local orientational order in hypoeutectic compositions, $\chi_B < 26\%$, as measured by the mean $\langle E \rangle$ of the extracopularity coefficient. The behavior of $\langle E \rangle$ versus $T$, depicted for cooling rate $R$ in \mbox{Fig. \ref{fig:statistics-mean}}, was observed to be qualitatively identical across all compositions and cooling rates, characterized by the strictly decreasing nature of $\langle E \rangle$ in both $\chi_B$ and $T$ and the lack of distinct features around any temperature. The slope of $\langle E \rangle$ was found to decrease with cooling rate.

Next, we studied the intensity of instantaneous fluctuations in orientational order for the same compositions, as measured by the variance $\mathrm{var}(E)$ of the extracopularity coefficient, whose behavior versus $T$ is depicted for cooling rate $R$ in Fig. \ref{fig:statistics-var}. Most distinctly, all hypoeutectic compositions were found to be approximate equal in $\mathrm{var}(E)$ at a certain temperature $T^*$, reported for our various cooling rates in Table \ref{tab:critical-values}. At the ensemble level, this phenomenon can be described through the rate of change in $\mathrm{var}(E)$ with respect to $\chi_B$ as follows:
\begin{equation}
    \left. \frac{\delta \mathrm{var}(E)}{\delta \chi_B} \right|_{T^*} \approx 0.
    \label{eq:composition-independence}
\end{equation} 
By contrast, the temperature behavior of the variance at $T^*$ was observed to be composition dependent. In particular, for $\chi_B > 15\%$, the the rate of change in $\mathrm{var}(E)$ with respect to $T$ at $T^*$ was positive, while for $\chi_B < 15\%$ it was negative. Neither result was observed to depend on cooling rate.

\begin{figure}[t]
    \centering
    \includegraphics[width=\linewidth]{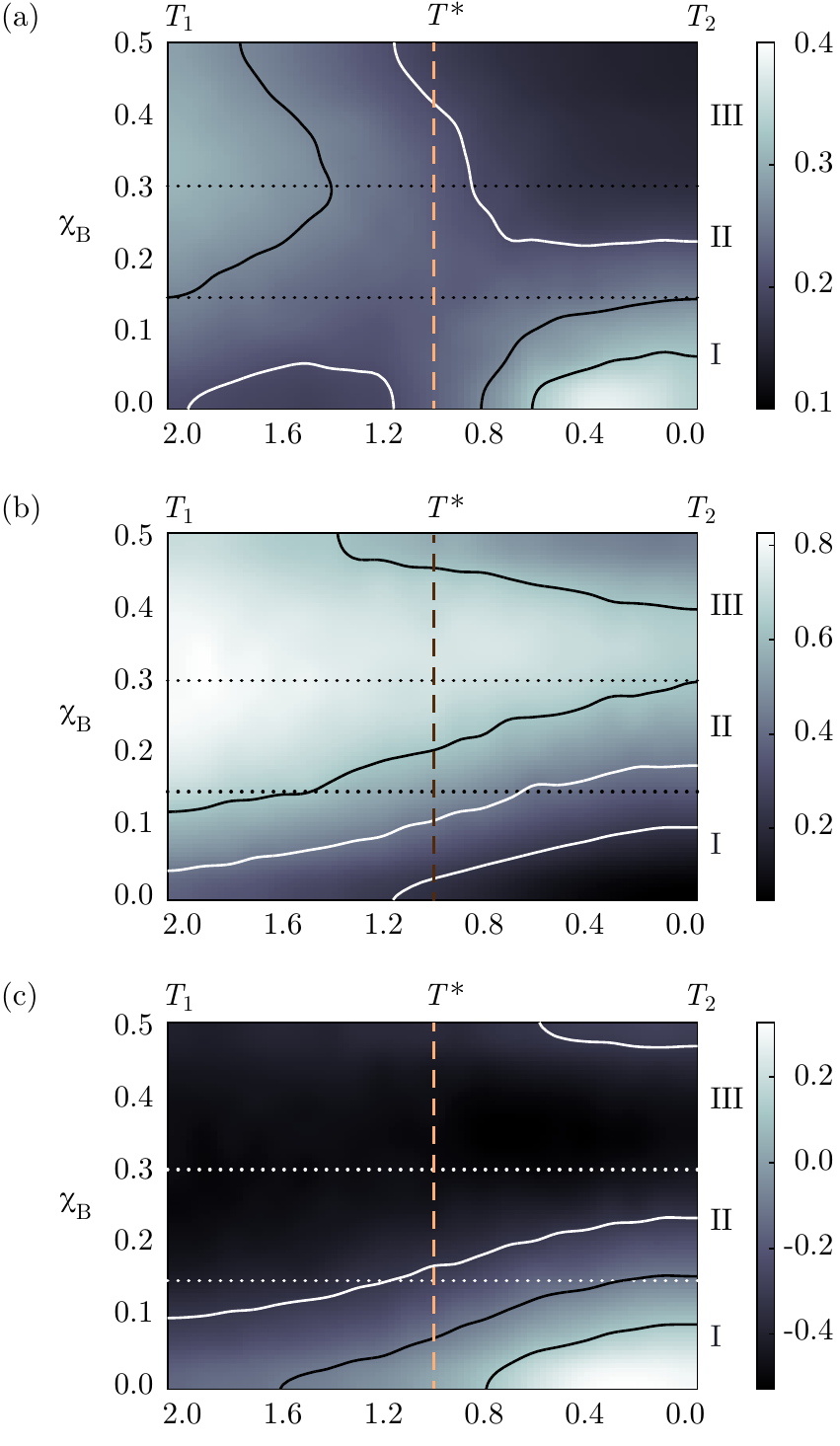} 
    \caption{Heatmaps of fluctuation intensity for KA-type mixtures cooled with rate $R$ in terms of $B$ molar fraction $\chi_B$ and temperature $T$: (a) LOF, (b) LDF, and (c) LSF. Roman numerals indicate the KA class corresponding to the regions demarcated by the dotted lines. The color gradients were obtained through the bicubic interpolation of our data.
    }
    \label{fig:heatmap}
    \makeatletter
    \let\save@currentlabel\@currentlabel
    \edef\@currentlabel{\save@currentlabel(a)}\label{fig:heatmap-lof}
    \edef\@currentlabel{\save@currentlabel(b)}\label{fig:heatmap-ldf}
    \edef\@currentlabel{\save@currentlabel(c)}\label{fig:heatmap-lrf}
\end{figure}

To get a more complete picture, we subsequently expanded the scope of our study to the full range of compositions, $0 \leq \chi_B < 50\%$. The mean $\langle E \rangle$ was found to sustain the trend seen in \mbox{Fig. \ref{fig:statistics-mean}}. Meanwhile, the behavior of $\mathrm{var}(E)$, depicted for rate $R$ in Fig. \ref{fig:heatmap-lof}, manifested a second criticality. In particular, while for $\chi_B < 30\%$ we observed $\mathrm{var}(E)$ to increase with $\chi_B$ for $T > T^*$, for $\chi_B > 30\%$ we observed $\mathrm{var}(E)$ to decrease with $\chi_B$ for those temperatures. \mbox{Fig. \ref{fig:heatmap}} compares the behavior of $\mathrm{var}(E)$ with its density and symmetry contributions, as defined in Eq. \ref{eq:decomposition}.
% EDIT LAST SENTENCE

Putting the two criticalities together, it can be said that KA-type mixtures were found to exhibit one of three qualitatively distinct behaviors, captured as \textit{classes} in the following piecewise expression:
\begin{equation}
    \mathrm{class}(\chi_B) = \! 
    \begin{cases}
        \text{I} &\! \text{if} \enspace \left. \dfrac{\delta \mathrm{var}(E)}{\delta \chi_B} \right|_{T > T^*} \!\! \geq 0, \left. \dfrac{\delta \mathrm{var}(E)}{\delta T} \right|_{T^*} \!\!< 0; \\[4mm]
        \text{II} &\! \text{if} \enspace \left. \dfrac{\delta \mathrm{var}(E)}{\delta \chi_B} \right|_{T > T^*} \!\! \geq 0,  \left. \dfrac{\delta \mathrm{var}(E)}{\delta T} \right|_{T^*} \!\! \geq 0; \\[4mm]
        \text{III} &\! \text{if} \enspace \left. \dfrac{\delta \mathrm{var}(E)}{\delta \chi_B} \right|_{T > T^*} \!\! < 0.
    \end{cases}
    \label{eq:piecewise}
\end{equation}

\paragraph{Analysis} The mean $\langle E \rangle$ of the extracopularity coefficient appears to behave in precisely the manner expected. In particular, the observation that $\langle E \rangle$ decreases with $\chi_B$ is consistent with the well-observed phenomenon that purer liquids tend to freeze into more ordered solids. Likewise, the observation that $\langle E \rangle$ decreases with temperature is consistent with the idea that increases in density imply higher levels of orientational order, though a general understanding of its exact form remains elusive \cite{leocmach_2012}. And the absence of discernible features in the behavior of $\langle E \rangle$ with respect to $T$ agrees with the traditional observation of essentially no change in structure over the glass transition \cite{guiselin_2022}. Thus, while it certainly serves to validate $E$ as an orientational order parameter, $\langle E \rangle$ does not by itself lead to much in the way of new insight.

By contrast, the behavior of $\mathrm{var}(E)$ suggests a highly informative, purely structural characterization of KA-type mixtures. This characterization---described mathematically by Eq. \ref{eq:piecewise} and summarized chemically in \mbox{Table \ref{tab:compositions}}---is in various ways consistent with the structural phase diagram obtained in Ref. \cite{chen_2018, pedersen_2018}. Notably, Class I mixtures correspond to those that rapidly crystallize into the FCC structure; Class II mixtures correspond to compositions associated with good glass-forming ability, including the standard ($80$:$20$) and eutectic ($74$:$26$) compositions; and Class III mixtures correspond to those whose equilibrium structure is CsCl. It is worth noting the breakdown in the composition independence observed near the point of equimolarity, possibly due to our cooling rates being too fast for the nucleation of crystals with CsCl structure. In summary, our observations on the variance of $E$ suggests an interesting correlation between the short-time changes in structure observed far from equilibrium and the thermodynamically stable structures that begin to manifests close to equilibrium.

\begin{table}[b!]
    \renewcommand{\arraystretch}{1.2}
    \centering
    \begin{ruledtabular}
        \begin{tabular}{lrrr}
            Cooling rate & Tdamp   & Pdamp  &  $T^*$   \\ \midrule 
            $R$          & $0.050$ & $1.00$ & $1.00$ \\
            $10R$        & $0.005$ & $1.00$ & $0.95$ \\
            $100R$       & $0.005$ & $0.05$ & $0.75$
        \end{tabular}
    \end{ruledtabular}
    \caption{For various cooling rates, our damping parameter settings and the observed temperature $T^*$ of composition independence.}
    \label{tab:critical-values}
\end{table}

\begin{table}[b!]
    \renewcommand{\arraystretch}{1.2}
    \centering
    \begin{ruledtabular}
        \begin{tabular}{llll}
            Composition & Class & Name                & Regime \\ \midrule 
            $100$:$0$   & I       & Pure              & FCC   \\
            $85$:$15$   & I       & Critically impure & FCC   \\
            $80$:$20$   & II      & Standard          & FCC   \\ 
            $74$:$26$   & II      & Eutectic          & PuBr$_3$ \\
            $70$:$30$   & III     & Superlative       & CsCl  \\
            $50$:$50$   & III     & Equimolar         & CsCl
        \end{tabular}
    \end{ruledtabular}
    \caption{Notable compositions for KA-type mixtures: the \textit{class} column indicates its qualitative behavior during rapid cooling (Eq. \ref{eq:piecewise}); the \textit{regime} column indicates its structure upon eventual crystallization \cite{chen_2018,pedersen_2018}.}
    \label{tab:compositions}
\end{table}

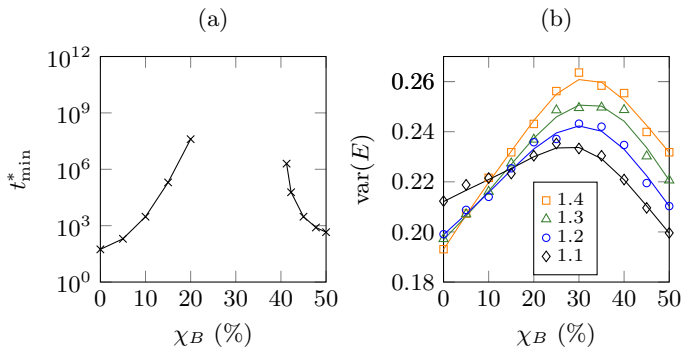
\begin{figure}[t]
    \begin{tabular}{lr}
    \hspace{7mm}
    \begin{tikzpicture}[baseline,trim axis left]
            \begin{semilogyaxis}[
                title={(a)},
                ylabel = $t^*_\mathrm{min}$,
                ylabel near ticks,
                ytick = {10^0,10^3,10^6,10^9,10^12},
                xtick={0,10,20,30,40,50},
                ymin=10^0,
                ymax=10^12,
                xlabel = {$\chi_B$ (\%)},
                xmin = 0,
                xmax = 50,
                width =0.53\linewidth,
                height=0.53\linewidth,
                /pgf/number format/.cd,
                1000 sep={},
                log basis y = 10,
                ]
                \addplot[black,mark=x,mark size=2pt] coordinates {
                (0,55)
                (5,200)
                (10,3000)
                (15,200000)
                (20,40000000)
                };
                \addplot[black,mark=x,mark size=2pt] coordinates {
                (41.25,2000000)
                (42.25,60000) % revise
                (45.00,3000) % revise
                (47.75,800) % revise
                (50.00,450)
                };
            \end{semilogyaxis}
        \end{tikzpicture}
        & \hspace{-3mm}
        \begin{tikzpicture}[baseline,trim axis right]
            \begin{axis}[
                title={(b)},
                ylabel = {$\mathrm{var}(E)$},
                ylabel near ticks,
                xlabel = {$\chi_B$ (\%)},
                ytick = {0.18,0.20,0.22,0.24,0.26,0.26},
                yticklabels = {0.18,0.20,0.22,0.24,0.26,0.26},
                legend style={font=\scriptsize,row sep=-0.1cm,/tikz/every odd column/.append style={column sep=0.01cm},at={(0.54,0.03)},anchor=south},
                legend cell align={left},
                xtick={0,10,20,30,40,50},
                xmin = 0,
                xmax = 50,
                ymin = 0.18,
                ymax = 0.27,
                width =0.53\linewidth,
                height=0.53\linewidth,
                ]
                % MARKS
                \addplot[orange,mark=square,mark size=1.4pt, only marks] coordinates {
                    (0,0.1931)
                    (5,0.2073)
                    (10,0.2216)
                    (15,0.2318)
                    (20,0.2431)
                    (25,0.2562)
                    (30,0.2636)
                    (35,0.2584)
                    (40,0.2554)
                    (45,0.2399)
                    (50,0.2318)
                };
                \addplot[OliveGreen,mark=triangle,mark size=2.0pt, only marks] coordinates {
                    (0,0.1972)
                    (5,0.2071)
                    (10,0.2161)
                    (15,0.2274)
                    (20,0.2369)
                    (25,0.2486)
                    (30,0.2495)
                    (35,0.2497)
                    (40,0.2487)
                    (45,0.2302)
                    (50,0.2206)
                };
                \addplot[blue,mark=o,mark size=1.4pt, only marks] coordinates {
                    (0,0.1991)
                    (5,0.2088)
                    (10,0.214)
                    (15,0.2254)
                    (20,0.2358)
                    (25,0.2371)
                    (30,0.2432)
                    (35,0.242)
                    (40,0.2347)
                    (45,0.2195)
                    (50,0.2103)
                };
                \addplot[black,mark=diamond,mark size=2.0pt, only marks] coordinates {
                    (0,0.2123)
                    (5,0.2188)
                    (10,0.2215)
                    (15,0.2233)
                    (20,0.2304)
                    (25,0.2353)
                    (30,0.2333)
                    (35,0.2304)
                    (40,0.2209)
                    (45,0.2096)
                    (50,0.1996)
                };
                % LINES
                \addplot[orange,no marks] coordinates {
                    (0,0.193)
                    (5,0.2067)
                    (10,0.2198)
                    (15,0.2322)
                    (20,0.2445)
                    (25,0.255)
                    (30,0.2608)
                    (35,0.2597)
                    (40,0.2526)
                    (45,0.2425)
                    (50,0.2317)
                };
                \addplot[OliveGreen,no marks] coordinates {
                    (0,0.1969)
                    (5,0.2063)
                    (10,0.2163)
                    (15,0.227)
                    (20,0.2374)
                    (25,0.2459)
                    (30,0.2506)
                    (35,0.2502)
                    (40,0.2442)
                    (45,0.2336)
                    (50,0.2206)
                };
                \addplot[blue,no marks] coordinates {
                    (0,0.1989)
                    (5,0.2071)
                    (10,0.2157)
                    (15,0.2247)
                    (20,0.2331)
                    (25,0.2395)
                    (30,0.2423)
                    (35,0.2402)
                    (40,0.233)
                    (45,0.2224)
                    (50,0.2103)
                };
                \addplot[black,no marks] coordinates {
                    (0,0.2121)
                    (5,0.2166)
                    (10,0.2209)
                    (15,0.2253)
                    (20,0.23)
                    (25,0.2335)
                    (30,0.2337)
                    (35,0.2292)
                    (40,0.2208)
                    (45,0.2104)
                    (50,0.1995)
                };
                \legend{1.4,1.3,1.2,1.1};
            \end{axis}
        \end{tikzpicture}
    \end{tabular}
    \caption{Measures of glass-forming ability: (a) the minimum time $t^*_\mathrm{min}$ needed for the crystallization of an $8000$ particle KA-type mixture of indicated composition \cite{pedersen_2018} and (b) the intensity of LOF for various temperatures above $T_\text{m}$, with fittings.}
    \label{fig:crystal-time}
\end{figure}

It is commonly suspected that the best KA-type glass former has a composition richer in the secondary component than the eutectic \cite{bruning_2008,pedersen_2018,ingebrigtsen_2019, mehri_2021}---the one traditionally associated with superlative glass-forming ability \cite{turnbull_1969}. The polynomial extrapolation of crystallization times performed by Pedersen and colleagues suggests that such a glass former would have a $B$ molar fraction of between $26\%$ and $36\%$ \cite{pedersen_2018}. Recent studies have found \textit{preordering} in liquids to be an important factor in determining where nucleation takes place upon sufficient cooling \cite{becker_2022, hu_2022b}. To the extent that $\mathrm{var}(E)$ captures the inhomogeneity of preordering in a liquid, it could constitute a purely structural heuristic for its glass-forming ability. Fig. \ref{fig:crystal-time} compares $\mathrm{var}(E)$ with crystallization time $t^*_\text{min}$, the traditional, dynamical measure of glass-forming ability. The visible correspondence between these two quantities suggests $70$:$30$, the composition ratio that appears to maximize $\mathrm{var}(E)$ in liquid state, to be that of the superlative KA-type glass former.

Our findings also appear to agree with the change in particle dynamics that is observed over the glass transition, namely the breakdown of Arrenhius temperature dependence. This change is often captured in an onset temperature $T_o$, though its widely varying estimates cast some doubt on its physical significance \cite{banerjee_2017}. We nevertheless note that the characteristic temperature at $10R$ coincides with the constant-pressure onset temperature $T_o = 0.95$ for $P=10\approx10.19$ reported by Coslovich and Pastore \cite{coslovich_2007}. Given that $T^*$ appears to be the temperature that most cleanly separates the two peaks in LOF, seen as \textit{hotspots} in Fig. \ref{fig:heatmap-lof}, understanding the structural significance of $T^*$ may hinge upon establishing the geometric origin of these features.

\begin{figure}[t]
    \begin{tabular}{l}
    \includegraphics[width=\linewidth]{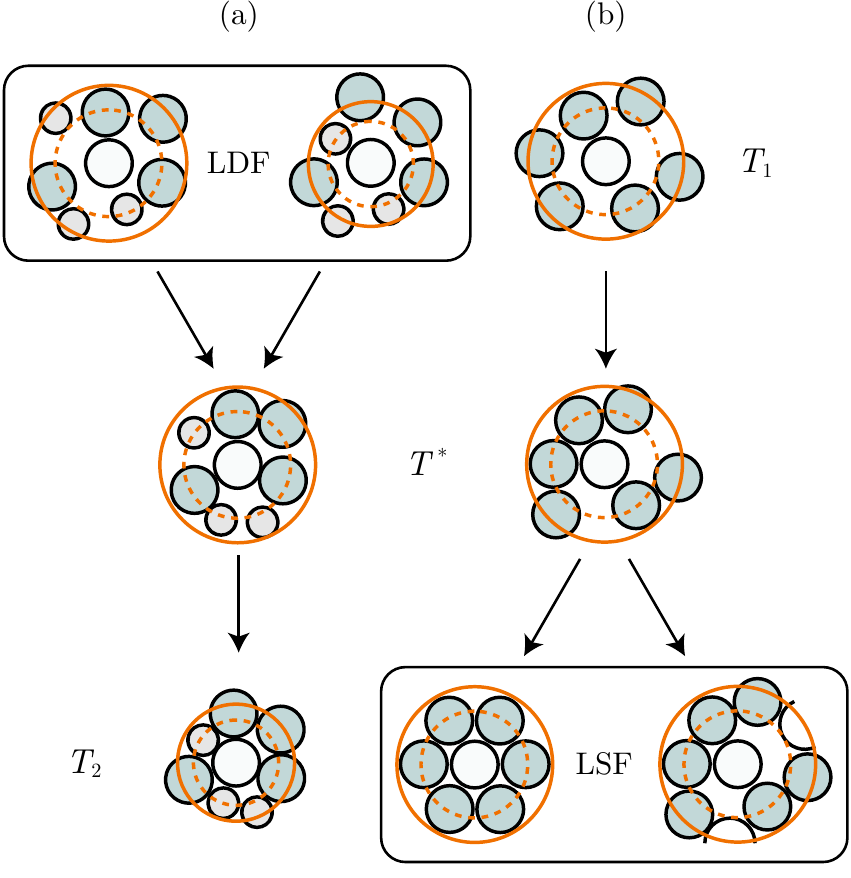}
    \end{tabular}
    \caption{A simplified, two-dimensional illustration of our explanation for the observed shift in the dominant source of LOF at $T^*$: (a) the decay in LDF upon the cooling of a binary liquid and (b) the growth in LSF upon the cooling of a pure liquid.
    }
    \label{fig:local-order-fluctuations}
    \makeatletter
    \let\save@currentlabel\@currentlabel
    \edef\@currentlabel{\save@currentlabel(a)(i)}\label{fig:local-order-fluctuations_a_i}
    \edef\@currentlabel{\save@currentlabel(a)(ii)}\label{fig:local-order-fluctuations_a_ii}
    \edef\@currentlabel{\save@currentlabel(b)}\label{fig:local-order-fluctuations_b}
\end{figure}

The larger hotspot is associated with high temperatures and impure mixtures, under which conditions the system constitutes a binary liquid. Therein, the dominant source of local-structural disparity are LDF, as visible from the comparison of Fig. \ref{fig:heatmap-ldf} and \ref{fig:heatmap-lrf}. LDF are observed to grow stronger with impurity concentration, likely on account of the possible difference in species between a particle and its nearest neighbor. Observe that an $A$ particle with nearest neighbor distance $r_{AA} = 2^{1/6} \sigma_{AA}$ will have a larger neighborhood than one with $r_{AB} < r_{AA}$. A larger neighborhood is able to capture more particles on average and therefore results in a higher expected extracopularity coefficient. This effect is illustrated schematically at the top of \mbox{Fig. \ref{fig:local-order-fluctuations_a_i}}.

The smaller hotspot corresponds to low temperatures and purer compositions, under which conditions the system constitutes an amorphous solid with a small impurity content. A comparison of Fig. \ref{fig:heatmap-ldf} with Fig. \ref{fig:heatmap-lrf} suggests the dominant source of local-order fluctuations to be LSF. This is consistent with the well-known phenomenon of geometric frustration \cite{crowther_2015}, whereby the tendency of one cluster toward the locally favorable configuration leaves adjacent clusters in a disfavorable, liquid-like configuration. The resulting disparity in local symmetry is illustrated schematically at the bottom of \mbox{Fig. \ref{fig:local-order-fluctuations_a_ii}}.

Now, the structural significance of the characteristic temperature $T^*$ can be understood by considering the interplay between the density and symmetry contributions to LOF. As defined in Eq. \ref{eq:composition-independence}, $T^*$ is the temperature at which small changes in composition lead to essentially no net change in LOF. So, at temperature $T^*$, changes with respect to composition in LSF must be equal in magnitude but opposite in direction to those in LDF. To the extent that LDF and LSF exemplify liquids and solids, respectively, $T^*$ constitutes the temperature at which the system is not made more or less solid-like by small changes in its composition. It can thus be interpreted as the temperature at which the structure of a system equally resembles that of a liquid and a solid.

\paragraph{Conclusion} The present work describes a novel statistical characterization of short-time structural relaxation in KA-type glass formers. The characterization is primarily facilitated by the second of the two statistics considered, namely the variance of the extracopularity coefficient $E$. This statistic measures the intensity of instantaneous fluctuations in orientational order and allows us to classify KA-type mixtures based on the way their structures relax when rapidly cooled. The intensity of these fluctuations also appears to constitute a heuristic measure of glass-forming ability, which could prove useful when exact methods are intractable. Perhaps most notably, this work establishes that the intensity of local-order fluctuations in a KA-type mixture is nearly independent of its composition at a temperature of putative dynamical significance. Our initial findings attribute this to the interaction between the density and symmetry contributions to instantaneous fluctuations in local orientational order. Further work is needed to establish the generality of this phenomenon and to understand the structure-dynamics relationship that may underlie it.

%TC:ignore

\bibliography{main.bib}

%TC:endignore

\end{document}